# Fourier-based classification of protein secondary structures

Jian-Jun SHU[*], Kian Yan YONG

*School of Mechanical & Aerospace Engineering, Nanyang Technological University, 50 Nanyang Avenue, Singapore 639798*

**ABSTRACT**

The correct prediction of protein secondary structures is one of the key issues in predicting the correct protein folded shape, which is used for determining gene function. Existing methods make use of amino acids properties as indices to classify protein secondary structures, but are faced with a significant number of misclassifications. The paper presents a technique for the classification of protein secondary structures based on protein "signal-plotting" and the use of the Fourier technique for digital signal processing. New indices are proposed to classify protein secondary structures by analyzing hydrophobicity profiles. The approach is simple and straightforward. Results show that the more types of protein secondary structures can be classified by means of these newly-proposed indices.

*Keywords*: Classification; protein secondary structures

## 1. Introduction

X-Ray crystallography and nuclear magnetic resonance spectroscopy are two widely-used instrumental methods [1] to determine protein secondary structures. Although these methods are powerful techniques for the structural determinations and analysis of proteins, they are resource-intensive and time-consuming. Because of these reasons, a bottleneck is created in the analysis of protein secondary structures, as protein database is growing exponentially in recent years. Hence computational methods are becoming useful in conjunction with instrumental methods in the prediction of protein secondary structures [2,3].

This paper deals with classifying protein secondary structures, namely $\alpha$-helix and $\beta$-strand, which are the building blocks of protein secondary structures. Several methods were proposed, including neural network [4] and wavelet transform [5]. Although these methods could predict the protein secondary structures with a reasonable level of accuracy, the types of protein secondary structures to be predicted were limited.

In this paper, a simple and effective novel method is developed to classify protein secondary structures, by utilizing the Fourier technique for digital signal processing. Unlike the wavelet transform and neural network techniques, this method does not require users to have the competency to select optimal parameters for each classification.

[*] Correspondence should be addressed to Jian-Jun SHU, mjjshu@ntu.edu.sg

A DNA sequence can be plotted as a signal by using a numerical representation of the four bases [6-9]. The same can be carried out for a protein sequence containing twenty types of amino acids, as shown in Fig. 1. Using this representation, DNA or protein sequences can be analyzed just like digital signal processing. Therefore the significant regions within DNA sequence, such as coding and non-coding regions [10,11], are expected to have different properties [12]. Similarly for protein secondary structures, such as $\alpha$-helix and $\beta$-strand, the different properties between them should be expected.

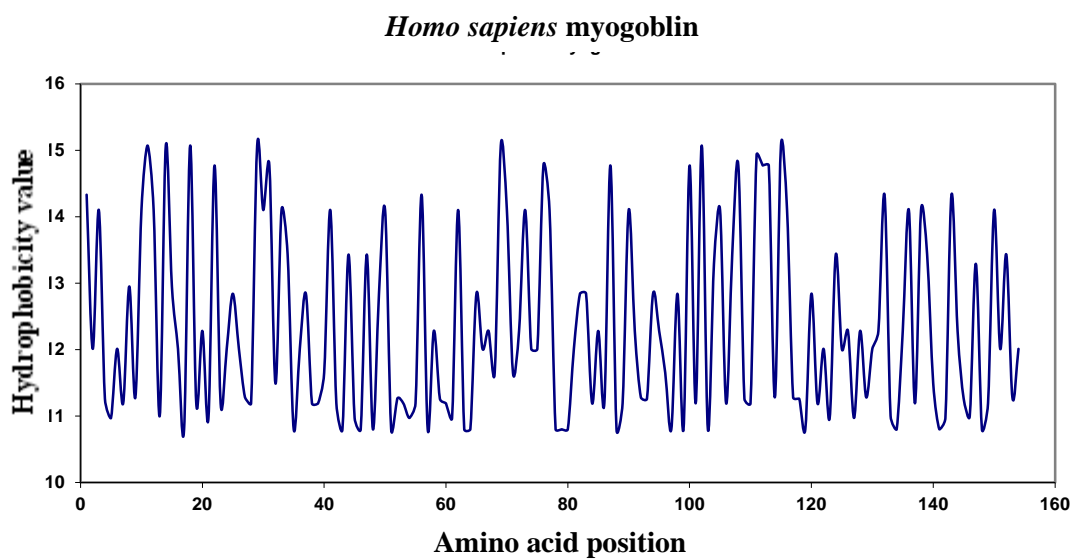

**Fig. 1.** Protein sequence plot of *Homo sapiens* myogloblin

Using these properties to analyze DNA and protein sequences opens up a novel field in sequence analysis in bioinformatics. In this paper, the properties, such as hydrophobicity and frequency, are used to classify protein secondary structures. A hypothesis based on the characteristics of protein secondary structures is drawn and verified based on existing experimental results.

## 2. Methods

The protein sequence of interest is obtained from GenBank in FASTA format. $\alpha$-helix and $\beta$-strand sequences are defined by twenty amino acids. There are many ways to encode amino acids numerically. For classifying protein secondary structures, hydrophobicity values are most relevant [13]. The hydrophobicity value $<H^f>$ is shown in Table 1 [14].

### Table 1

Bulk hydrophobic character and accessibility coefficients for *20* amino acids

| Amino acid | $<H^f>$ | $<Br>$ |
|---|---|---|
| Ala | 12.28 | 0.39 |
| Arg | 11.49 | 0.12 |
| Asn | 11.00 | 0.16 |
| Asp | 10.97 | 0.17 |
| Cys | 14.93 | 0.58 |
| Gln | 11.28 | 0.16 |
| Glu | 11.19 | 0.16 |
| Gly | 12.01 | 0.38 |
| His | 12.84 | 0.29 |
| Ile | 14.77 | 0.55 |
| Leu | 14.10 | 0.48 |
| Lys | 10.80 | 0.13 |
| Met | 14.33 | 0.46 |
| Phe | 13.43 | 0.35 |
| Pro | 11.19 | 0.22 |
| Ser | 11.26 | 0.25 |
| Thr | 11.65 | 0.26 |
| Trp | 12.95 | 0.23 |
| Tyr | 13.29 | 0.30 |
| Val | 15.07 | 0.63 |

From previous work [14], the profiles of protein secondary structures for exposed helical structure, exposed $\beta$-structure, $\beta$-turn and buried $\beta$-structure are produced and classified. In Fig. 2, exposed helical and exposed $\beta$-structures are easily distinguishable from $\beta$-turn and buried $\beta$-structure. The former group has a shape that approximates a sine wave whereas the latter group composes of a U shape and an inverted U shape curve respectively. By using the mean difference values of each point from the critical hydrophobicity value of *12*, $\beta$-turns can be identified, because the amino acids of $\beta$-structures have the low affinity with water. For the classification of exposed helical and exposed $\beta$-structures, the hydrophobicity profiles are used.

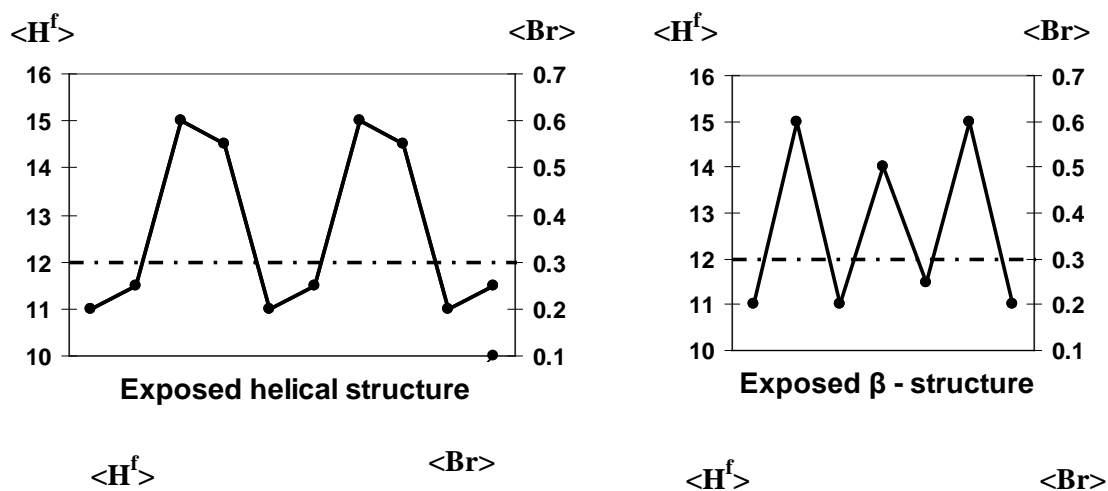

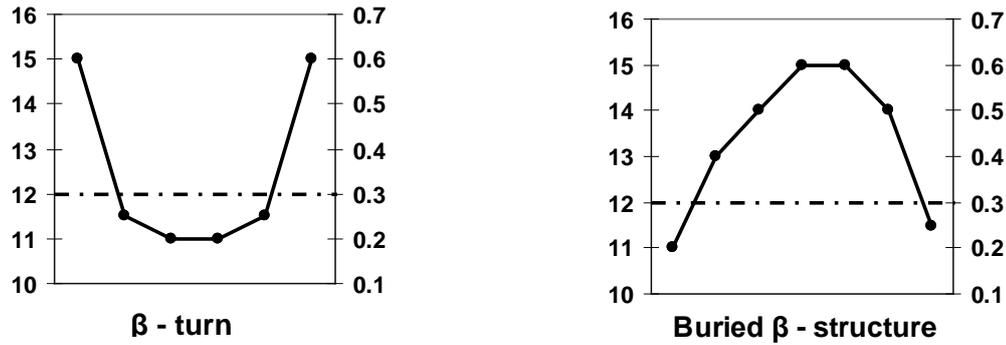

**Fig. 2.** *4* basic hydrophobicity profiles; <H$^f$>, <Br> versus base positions [14]

## 3. Implementation

**Classification of bovine phospholipase A$_2$**

The classification of protein secondary structures is implemented on a set of known protein secondary structures shown in Table 2 [14]. The protein bovine phospholipase A$_2$ is used for demonstration. According to experimental data, bovine phospholipase A$_2$ contains *5* $\alpha$-helices, *3* exposed $\beta$-sheets and *6* $\beta$-turns. Here it is shown how the *3* groups are classified using a two-dimensional classification plot. The bovine phospholipase A$_2$ protein sequence is taken from GenBank under the accession code of 1KVX. From there, the individual known substrings are gathered. Convert protein sequence into a signal with amplitude and time. The amplitude of each amino acid is assigned using the hydrophobicity value <H$^f$> from Table 1. Each unit of time corresponds to one amino acid. A signal graph of *Homo sapiens* myoglobin is shown in Fig. 1. Break up the signal into its frequency components. This is done by using the Fourier transform as follows

$$X_k = \frac{1}{N}\sum_{n=0}^{N-1} x_n e^{i2\pi k \frac{n}{N}}$$

where *N* is the number of amino acids and each substring $x_n$ is numerically represented. Amplitudes in the frequency domain are henceforth called hydrophobicity value $X_k$. Dominant frequency *k* corresponds to the one with the highest hydrophobicity value. The classification plots of the protein sequence 1KVX are generated in Fig. 3.

**Table 2**

Set of protein sequences with known protein secondary structures [14]

| Protein | AAR | Resolution (Å) | Q$_\alpha$(%)[b] | Q$_\beta$(%)[c] |
|---|---|---|---|---|
| Adenylate kinase | 194 | 3.0 | 82.1 | 97.2 |
| Alcohol dehydrogenase[a] | 374 | 2.4 | 67.5 | 73.5 |
| Carbonic anhydrase C | 259 | 2.0 | 82.5 | 80.0 |
| Carboxypeptidase A | 307 | 2.0 | 67.0 | 80.4 |
| Catalase | 505 | 2.5 | 70.5 | 76.5 |
| Concanavalin A[a] | 238 | 2.4 | 96.6 | 74.3 |
| Cytochrome b$_s$ | 93 | 2.8 | 75.6 | 82.3 |
| Dihydrofolate reductase | 189 | 2.9 | 86.4 | 76.6 |
| Ferricytochrome c | 128 | 2.5 | 76.5 | 95.0 |
| Flavodoxin | 138 | 1.8 | 93.4 | 88.5 |
| Lamprey globin | 148 | 2.0 | 71.3 | 85.8 |
| Glutathione reductase[a] | 478 | 2.0 | 69.8 | 69.9 |
| D-Glyceraldchyde-3-phosphate dehydrogenase[a] | 334 | 2.9 | 79.7 | 77.3 |
| Lactate dehydrogenase domain I | 178 | 2.0 | 80.5 | 85.9 |
| Lysozome | 129 | 1.5 | 83.1 | 66.2 |
| Phospholipase A$_2$ | 122 | 1.7 | 90.5 | 89.0 |
| Rhodanase domain I[a] | 137 | 2.5 | 89.6 | 88.2 |
| Ribonuclease | 124 | 2.0 | 79.4 | 75.2 |
| Subtilisin | 275 | 2.5 | 85.0 | 87.7 |
| Cu, Zn Superoxide dismutase | 151 | 3.0 | 91.4 | 80.7 |
| Thermolysin[a] | 316 | 2.3 | 73.6 | 67.3 |
| Thioredoxin-S$_2$[a] | 108 | 2.8 | 70.5 | 86.2 |
| Triose phosphate isomerase | 248 | 2.5 | 70.5 | 86.6 |

[a] Some reported $\beta$-structures have < 5 AAR and/or helices with < 6 AAR
[b] Average $Q_\alpha = 79.9\%$ ; [c] Average $Q_\beta = 80.7\%$

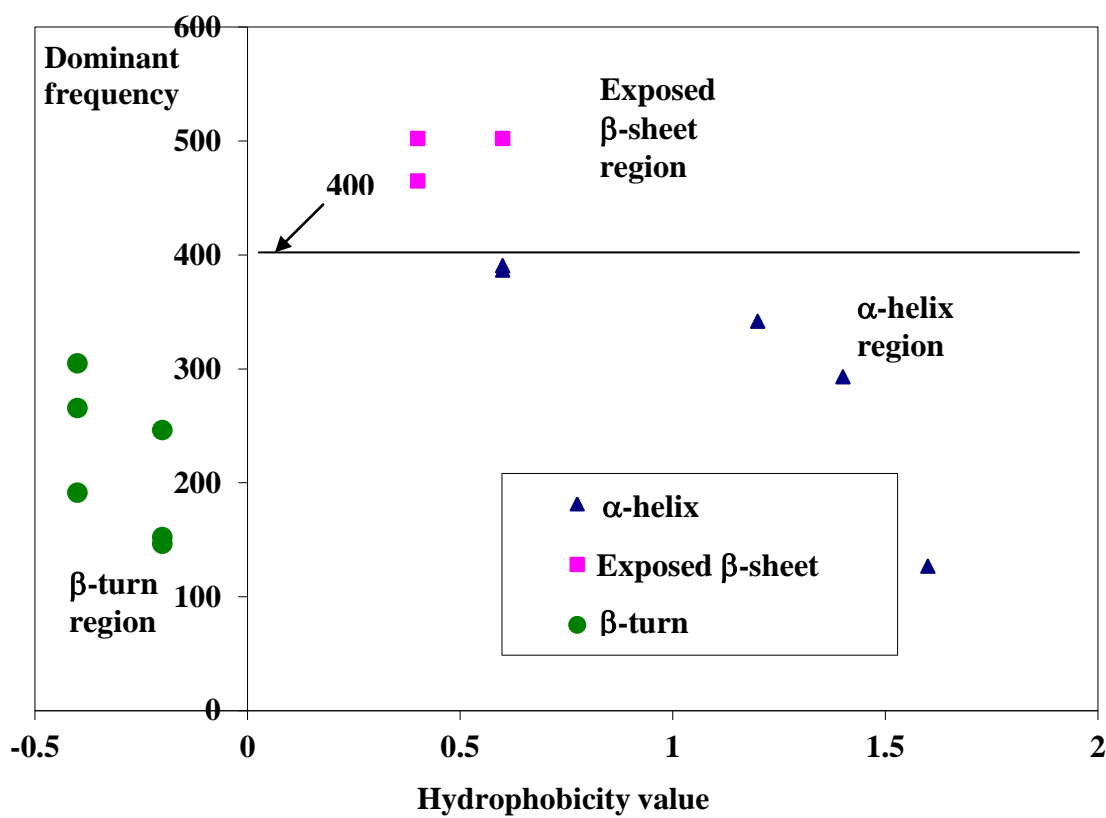

**Fig. 3.** Classification plot of hydrophobicity value versus dominant frequency for 1KVX

## 4. Discussion

It is evident that from hydrophobicity value alone, $\beta$-turns are classified in the negative $x$-axis due to the majority of amino acids being less hydrophobic (below critical hydrophobic value of *12*). The other type of protein secondary structures has positive hydrophobicity value and falls on the right side of the origin. This can be seen from Fig. 3. In order to differentiate the remaining types of protein secondary structures – $\alpha$-helix and exposed $\beta$-sheet – dominant frequency is used. Since the amino acids of exposed $\beta$-sheets alternates between positive and negative hydrophobicity values at a greater frequency than that of $\alpha$-helix structure, the plot of the former is expected to be higher in the graph. This point could well be the buried helical structure of similar shape to the buried $\beta$-structure, which is not defined by [14] in Fig. 2. From the above analysis, a hypothesis can be derived for the family of the bovine phospholipase proteins as summarized in Table 3.

**Table 3**

Hypothetical classifications of three protein secondary structures

|  |  | **Dominant frequency** | |
|---|---|---|---|
|  |  | **< 400** | **> 400** |
| **Hydrophobicity value** | **< 0** | $\beta$-**turn** | |
|  | **0 ~ 0.55** | **Buried** $\beta$-**sheet** | **Exposed** $\beta$-**sheet** |
|  | **> 0.55** | $\alpha$-**helix** | |

More tests on other bovine phospholipase proteins may be required to prove the above hypothesis as well as the critical dominant frequency value to distinguish between exposed $\alpha$-helix and $\beta$-sheet. Experimental data on exposed and buried $\alpha$-helix are useful in the understanding of how amino acids fold under different scenarios. Besides bovine phospholipase proteins, other proteins can be tested using these indices in order to come up with a universal tool to distinguish protein secondary structures.

A further test is carried out on selected three $\alpha$-helix structures from adenylate kinase (GenBank 3ADK) and a total of four buried and exposed $\beta$-structures from concanavalin A (GenBank P81461) [14]. The result is shown in Fig. 4. From the hypothesis of Table 3, *2* out of *3* $\alpha$-helix structures are correctly classified. Buried $\beta$-sheets are generally plotted below exposed $\beta$-sheets and on the left of $\alpha$-helices.

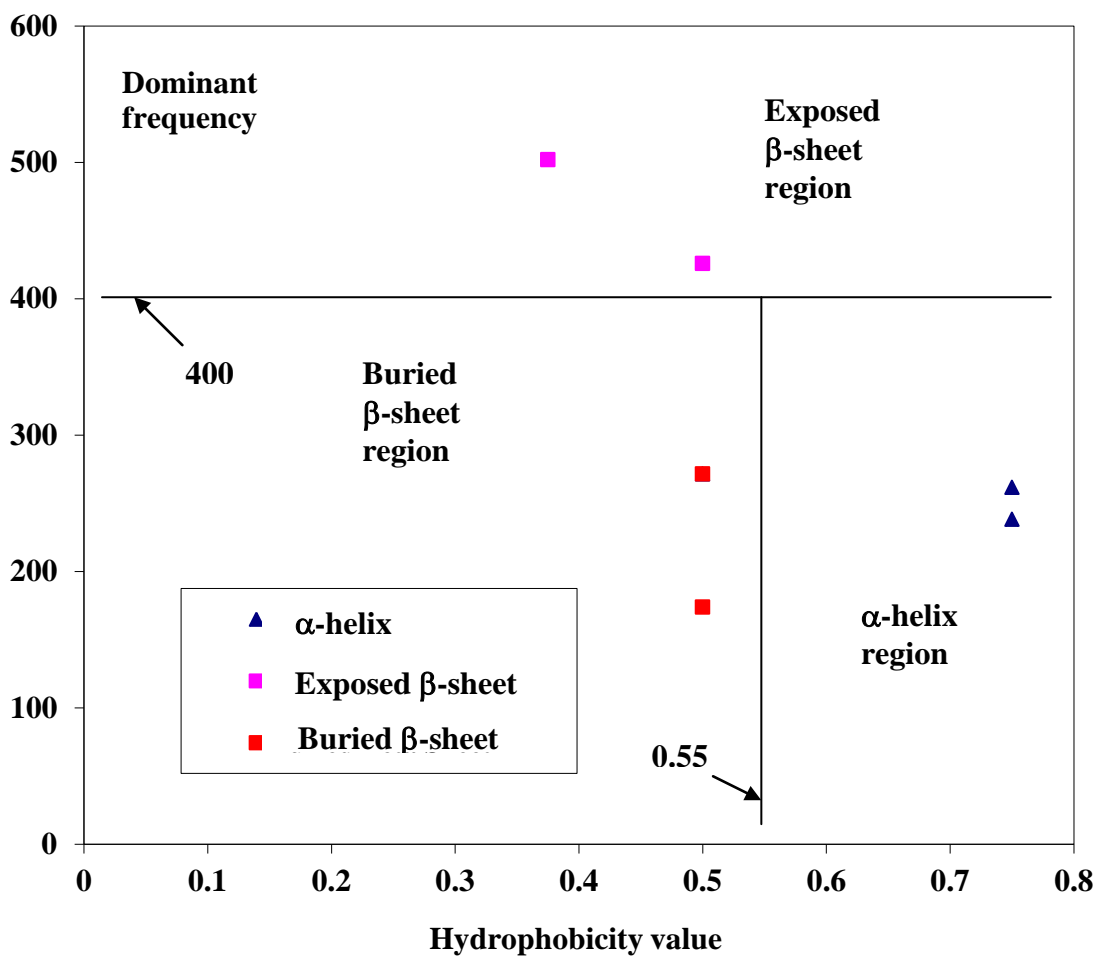

**Fig. 4.** Classification plot of selected protein secondary structures from 3ADK and P81461

It is worth noting that based on Fig. 4, buried $\beta$-sheet, which cannot be detected [14], is shown to be in a distinguishable group. The $\alpha$-helix with a high hydrophobicity value of greater than *0.55* suggests that many of its amino acids are hydrophilic. Hence they can fold in such a way that these amino acids do not come into contact with water under normal circumstances, thus preventing a reaction from taking place. In order for that to happen, the amino acids need to be buried within the protein. This means that the $\alpha$-helix can in fact be buried in nature, giving rise to another protein secondary structure that is not classified [14] in Fig. 2. Evidently from Fig. 4 alone, it shows that the more types of protein secondary structures can be classified using this new set of indices, at least within the 3ADK and P81461 class of protein sequences.

In order to verify the hypothesis set in Table 3, Figs. 3 and 4 are combined into Fig. 5, which shows that all protein secondary structures are well classified. $\beta$-turns are clearly situated in the negative hydrophobicity value region on the left hand side of the graph. This implies that they are hydrophobic in nature. Both $\alpha$-helices and $\beta$-sheets are situated in the positive hydrophobicity value region. $\alpha$-helices are generally situated in the region with a dominant frequency value of less than *400*. This implies that there is less volatility or differences among the hydrophobicity values of amino acids to form structures. As for $\beta$-sheets, the majority of them are situated above the *400* line. The hypothesis is verified.

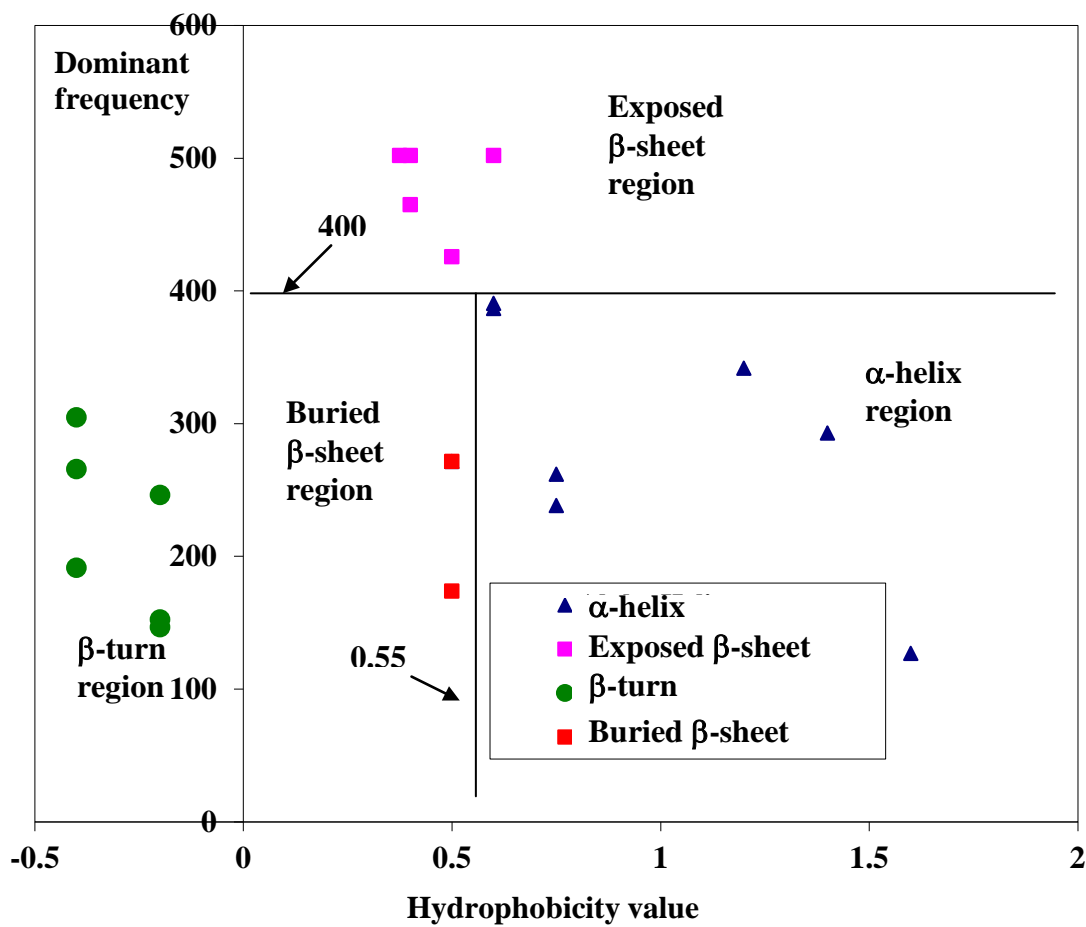

**Fig. 5.** Classification plot of 1KVX, 3ADK and P81461

## 5. Conclusion

In this paper, new indices have been proposed to classify protein secondary structures. It has been shown that using these indices – hydrophobicity value and dominant frequency – it is possible to classify the more types of protein secondary structures. Ultimately, it is hope that this research opens up a whole new concept for the analysis of not only protein secondary structures, but also DNA and protein sequences [15-18].

**Acknowledgements**

This work was supported by Nanyang Technological University (M4081942).